# High-Mobility Carriers in Epitaxial IrO$_2$ Films Grown using Hybrid Molecular Beam Epitaxy

Sreejith Nair[1,*], Zhifei Yang[1,2], Kevin Storr[3] and Bharat Jalan[1,*]

1. Department of Chemical Engineering and Materials Science, University of Minnesota, Twin Cities, Minnesota, 55455, USA

2. School of Physics and Astronomy, University of Minnesota, Twin Cities, Minnesota, 55455, USA

3. Department of Physics, Prairie View A&M University, Texas, 77446-0519, USA

[*]Corresponding authors: nair0074@umn.edu; bjalan@umn.edu

**Abstract:**

Binary rutile oxides of 5*d* metals such as $IrO_2$, stand out as a paradox due to limited experimental studies despite the rich predicted quantum phenomena. Here, we investigate the electrical transport properties of $IrO_2$ by engineering epitaxial thin films grown via hybrid molecular beam epitaxy. Our findings reveal phonon-limited carrier transport and thickness-dependent anisotropic in-plane resistance in $IrO_2$ (110) films, the latter suggesting a complex relationship between strain relaxation and orbital hybridization. Magneto-transport measurements reveal a previously unobserved non-linear Hall effect. A two-carrier analysis of this effect shows the presence of minority carriers with mobility exceeding 3000 $cm^2$/Vs at 1.8 K. These results point towards emergent properties in 5*d* metal oxides that can be controlled using dimensionality and epitaxial strain.

**Introduction:**

The rutile crystal structure has emerged over the years, as a host of exotic quantum effects with the discovery of unconventional properties in materials like $VO_2$ and $RuO_2$. *3d* $VO_2$ has now been studied for decades due to its near room temperature coupled structural and metal-insulator transition[1, 2] for use in smart windows[3], field effect Mott transistors[4] and neuromorphic computing[5]. *4d* $RuO_2$ on the other hand, has come to the forefront in condensed matter physics more recently due to the discovery of strain-induced superconductivity[6], non-trivial band topology[7], altermagnetism[8-11] - a novel magnetic state in condensed matter, and associated spintronic functionalities[12-14]. However, the *5d* counterpart $IrO_2$, although of great technological interest, has received relatively less attention.

The interest in $IrO_2$[15] and the iridate family in general, has multiple origins and motivations. Countless predictions of non-trivial topology[16, 17], quantum criticality[18] and unconventional superconductivity[19] have surrounded the complex iridium oxides. One of the key examples for instance, has been the prediction and experimental observation of topological line nodes[20, 21] and the large Berry curvature induced spin-Hall conductivity in perovskite $SrIrO_3$[22]. Such large spin-Hall conductivities can significantly improve device performance with the perovskite structure allowing further integration with a host of other material systems. However, from a commercial standpoint, binary oxides may be preferred due to ease of synthesis and reproducibility. With growing literature on perovskite-like quantum properties in simple binary oxides, they can form a comprehensive quantum system that is efficient and scalable.

Like the perovskite $SrIrO_3$, rutile $IrO_2$ has been shown to have Dirac nodal lines (DNLs) close to the Fermi level using density functional theory[23] (DFT) and angle-resolved photoemission spectroscopy (ARPES) measurements[24-26]. There have also been reports on high spin-Hall conductivity arising due to these DNLs[27, 28] in line with theoretical predictions[23]. A significant amount of work has also been done to confirm the presence of the predicted $J_{eff}$ =1/2 states[29-31] using spectroscopic techniques, although the debate remains unsettled. In contrast to the numerous spectroscopic characterizations, only a few studies[32-34] have reported the normal state electronic transport in rutile $IrO_2$, particularly in epitaxial thin film form.

The rutile crystal structure (hereafter described using the tetragonal *a=b* and *c* axes) inherently lends many intriguing aspects to the electronic structure. The mixed edge and corner shared anion octahedra as shown in Fig. 1a and 1c, is a key ingredient for most unconventional properties in rutile systems[35] with the metal-metal dimerization along the $c$-axis in $VO_2$ being one of the prime examples. Further, the low symmetry ($D_{2h}$) distorted octahedral ligand field, creates a non-degenerate $t_{2g}$ band with the $d$ orbital parallel (denoted $t_{\parallel}$ in the Goodenough notation[36]) to the $c$-axis ($d_{x^2-y^2}$ orbital in Fig. 1c) having minimal hybridization with O $2p$ orbitals and density of states at the Fermi level[29, 30, 37].

With this context of topology and orbital character of the Fermi surface, we analyzed the electronic transport in epitaxial $IrO_2$ thin films measured as a function of temperature, magnetic field, dimensionality, strain, and epitaxial orientation. The temperature-dependent resistivity from 300K to 1.8 K revealed a phonon dominated carrier scattering with resistivity temperature exponent $n > 1$, consistent with a Debye temperature higher than 300 K, extracted from a Bloch-Grüneisen type model. Further, in agreement with bulk $IrO_2$[38], the resistance along rutile $c$-axis was found to be greater than the perpendicular directions (i.e. along a- and b-axis). This anisotropy was also found to be thickness-dependent, suggesting a role of epitaxial strain on the Ir $5d$ - O $2p$ hybridization. Hall effect measurements showed a non-linear dependence of Hall resistance on magnetic field. A two-carrier analysis revealed the presence of high-mobility minority carriers. The validity of the two-carrier model was further confirmed with the observation of electron or hole dominated conduction in $IrO_2$ films with varying epitaxial orientations. This emergence of previously unobserved high mobility carriers in $IrO_2$ opens new questions and avenues for engineering quantum phases in iridium-based oxides by tuning the interplay of strain, dimensionality, and electronic structure.

**Results and Discussion:**

Using the hybrid MBE approach[39], we grew single-crystalline, epitaxial $IrO_2$ films with varying thicknesses and orientations on $TiO_2$ substrates. The details of growth conditions, sample characterization and strain relaxation can be found elsewhere[40], and are briefly described in the Methods section of the *Supplementary Information*. Fig. 2a shows the temperature-dependent resistivity of $IrO_2$ (110) films for 2.5 nm $\leq t \leq$ 21 nm measured in the Van der Pauw configuration. The in-plane resistance anisotropy, defined as the ratio of resistance along [001] and [1$\bar{1}$0] directions i.e. $R_{[001]}/R_{[1\bar{1}0]}$ at 300 K is shown as a function of film thickness in the inset of Fig. 2a.

The resistance anisotropy ranges between 2-5 with a non-monotonic behavior with increasing thickness.

We first discuss the temperature-dependent resistivity followed by discussion of the resistance anisotropy. As shown in Fig. 2a, metallic behavior was observed in films with thickness as low as 2.5 nm. A clear upturn in resistivity was observed around 50 K for the 2.5 nm film, suggesting a localization induced upturn. Such a resistance upturn with decreasing thickness has been observed in many systems[41-43] including $IrO_2$[32] and the mechanisms have been well documented. However, to understand the different scattering contributions to the resistivity, we analyzed our *T*-dependent resistivity data using a Bloch-Grüneisen type model (see *Supplementary Information* for more details). Equation 1 which is a combination of the residual resistivity ($\rho_o$), resistivity due to electron-electron scattering ($\rho_{ee}$), acoustic ($\rho_{e-ph,ac}$) and optical ($\rho_{e-ph,op}$) phonon scattering, was used to fit the temperature-dependent electrical resistivity.

$$\rho(T) = \rho_o + \rho_{ee}(T) + \rho_{e-ph,ac}(T) + \rho_{e-ph,op}(T)$$

$$= \rho_o + A_{ee}T^2 + \beta_{ac}T\left(\frac{T}{\theta_D}\right)^2 \int_0^{\theta_D/T} \frac{x^3}{(e^x-1)(1-e^{-x})} dx + \beta_{op}T\left[\frac{\theta_E/2T}{\sinh(\theta_E/2T)}\right]^2 \quad (1)$$

$A_{ee}$ is the prefactor for the electron-electron scattering term, $\beta_{ac}$ and $\beta_{op}$ are a measure of the electron-phonon coupling strength for the acoustic and optical branches, $\theta_D$ and $\theta_E$ represent the Debye temperature and the temperature corresponding to the Einstein frequency of the optical phonon spectrum respectively.

Fig. 2b shows excellent fits to the experimental data using equation 1 with $\rho_o$, $A_{ee}$, $\beta_{ac}$, $\beta_{op}$, $\theta_D$ and $\theta_E$ as the fit parameters. For brevity, only a few thicknesses are shown, and the 2.5 nm was excluded from these analyses due to the low-temperature resistance upturn. The extracted fit parameters for the different film thicknesses are tabulated in Table S1. The extracted constants for the electron-phonon coupling contributions are close to previous reports[44, 45]. Their ratio $\beta_{ac}/\beta_{op}$ is also in agreement with the calculated values of the closely related rutile $RuO_2$ (~ 2) by Glassford and Chelikowsky[46]. The electron-electron scattering term, although included here for better fits at temperatures less than 50-60 K, can be neglected without significant differences in the extracted fit parameters and regression coefficient for the overall fit as shown in Table S2. The Debye temperature in both cases is around 430-460 K which is higher than that reported by Lin et al[44], but much lower than the value obtained by Ryden et al[38]. Nevertheless, the extracted Debye temperature can be verified based on the expectation of linear $T$-dependence of resistivity above the Debye temperature for normal metals. Or in other words, the first order derivative of resistivity with respect to temperature should approach a constant value with increasing temperature above $\theta_D$ (430 - 460 K in this case). To this end, we show in Fig. 2c, the first derivative of resistivity ($d\rho/dT$) as a function of $T$ between 50 K and 300 K. The inset shows $d\rho/dT$ vs $T$ data for a representative 15 nm $IrO_2/TiO_2$ (110) up to 390 K. This result shows a nearly constant $d\rho/dT$ value as $T$ increases which is reasonably consistent with our extracted values of $\theta_D$ = 430 - 460 K.

It is important to note here that the resistivity analysis was performed on an averaged resistivity[47] obtained using the Van der Pauw equation. The measured resistances along the two orthogonal in-plane crystallographic directions as mentioned earlier, are however anisotropic, similar to the observation in bulk single crystals where the ratio of resistivity along the $c$ and $a$ axis ($\rho_c/\rho_a$) is approximately 1.8[38]. As shown in the inset of Fig. 2a, the anisotropy ($R_{[001]}/R_{[1\bar{1}0]}$) ranges between 2-5 in our $IrO_2$ thin films, initially increasing and then decreasing as $t$ increases. This observation raises questions about the origin of anisotropy and why it depends on film thickness. We attribute the former to the anisotropic crystal structure along the [001] and [1$\bar{1}$0] directions, and the latter to the strain relaxation in epitaxial $IrO_2$ (110) films. We attempt to explain the origin of anisotropy qualitatively by referring to the schematics in Figs. 1a and 1c. As described briefly in the introduction, the rutile structure has a network of edge-shared octahedra along the $c$-axis and corner-shared octahedra along the direction perpendicular to the $c$-axis. This arrangement is expected to yield contributions to the electrical conduction from the metal-oxygen (M $d$- O $p$) orbital hybridization of the M-O-M network and direct M-M hopping along the $c$-axis, whereas the contribution is only from the M-O-M network along the direction perpendicular to $c$-axis. This effect of crystal anisotropy and the relative M-M and M-O-M transport contributions to the resistivity anisotropy at a cursory glance, seems to be captured by the tetragonality ($c/a$) of the rutile crystal. As shown in Fig. 1b, there appears to be a correlation between the reported resistivity ratio ($\rho_c/\rho_a$) and the $c/a$ lattice parameter ratio in bulk metallic rutile oxides.

However, when it comes to the thickness dependence of anisotropy, there is an additional effect of epitaxial strain relaxation. Epitaxial $IrO_2$ on $TiO_2$ (110) substrate experiences a large anisotropic strain with a -6.3 % compressive strain along the [001] direction and a +2.3 % tensile strain along the [1$\bar{1}$0] direction. So, for a coherently strained film, $c/a$, and hence, the resistivity anisotropy would be expected to be the lowest and eventually increase towards the bulk value with increasing thickness and associated strain relaxation. This expectation however, contrasts with the observed non-monotonic variation in anisotropy with thickness shown in the inset of Fig. 2a. This necessitates a more detailed consideration of the intricacies of the anisotropic strain induced octahedral distortion and relative transport contributions from the M-M and M-O-M channels.

For the case of IrO$_2$, since the $t_{\parallel}$ orbital ($d_{x^2-y^2}$ orbital in Fig. 1c which governs Ir-Ir interaction along $c$-axis) has been shown to have minimal contribution to the density of states at Fermi level[29,30], one can expect negligible transport contribution from direct Ir-Ir hopping, assuming the absence of a significant strain induced shift of the $t_{\parallel}$ density of states towards the Fermi level. So, the anisotropy in resistance should largely depend on the Ir-O-Ir network along the two in-plane directions, with larger bond angles favoring Ir 5$d$ - O 2$p$ hybridization and resulting in decreased resistance. For a relaxed bulk-like IrO$_2$ film, the bond angle is higher along [1$\bar{1}$0] (~ 128º) as compared to the [001] direction (~ 103º) and hence, R$_{[1\bar{1}0]}$ can be lower than R$_{[001]}$. However, this difference in bond angles and resulting resistance anisotropy can be amplified by the anisotropic strain in IrO$_2$ (110) film grown on TiO$_2$ (110) substrate. The compressive strain along [001] (-6.3%) will push the Ir-O-Ir angle to smaller values (diminishing $d$-$p$ hybridization) whereas the tensile strain along [1$\bar{1}$0] (+2.3%) will drive the Ir-O-Ir angle to larger values closer to 180º (enhancing $d$-$p$ hybridization). This change in bond angles, however, is also accompanied by a change in bond lengths. The compressive strain along [001] will decrease the Ir-O bond length (enhancing $d$-$p$ hybridization) whereas the tensile strain along [1$\bar{1}$0] will lead to an increase in the bond length (diminishing $d$-$p$ hybridization). From this simplified picture of thickness-driven strain relaxation and changes in the Ir-O bond network, the evolution of bond angles towards bulk values with increasing thickness should result in a monotonic *decrease* in the anisotropy (R$_{[001]}$/R$_{[1\bar{1}0]}$) with increasing thickness. Whereas the simultaneous change in bond lengths should lead to a monotonic *increase* in the anisotropy. Thus, the variation in bond lengths and bond angles as a function of thickness and strain relaxation may act as competing effects and the interplay of the two could lead to the observed non-monotonic variation of the resistance anisotropy. A more careful quantitative study of crystal defects, octahedral distortions, orbital hybridizations and resulting carrier hopping integrals as a function of film thickness and associated strain is however required, to accurately predict and engineer anisotropy for different electronic and spintronic applications. It is also equally important to note that the effect of strain relaxation is coupled with changing film thickness which as a dimensionality effect, can change the electronic structure, especially in systems like $IrO_2$ with non-trivial topology and possible surface states[25].

To further probe the effect of thickness on the electrical transport, we performed Hall effect measurements for the $IrO_2$ (110) films. Fig. 3a shows the Hall conductance at 1.8 K for different film thicknesses, as defined in equation 5 of the *Supplementary Information*, plotted against the applied magnetic field. We observed dominant hole carriers with increasing concentration (increasing slope) as a function of film thickness. A closer look at the Hall conductance revealed a sizable non-linearity for the low film thicknesses as shown in the inset of Fig. 3a (Hall conductance for the 5.5 nm film). The thickness dependence of non-linear Hall effect can be better visualized using the first order derivative of Hall conductance plotted as a function of applied magnetic field. As shown in Fig. S2a, a peak is observed in the derivative plot near zero magnetic field, a feature that vanishes with increasing film thickness. The non-linearity can also be estimated using the difference between the experimental data and linear fits to the low-field region as shown in Fig S2b. The separation between the experimental data and linear fit can be observed to decrease with increasing film thickness. This is quantitatively captured in Fig. S2c, where the percentage difference between the linear fit and the experimental Hall conductance at 9T is plotted as a function of film thickness. An exponential-like decay was observed for this parameter with increasing thickness. The non-linear Hall effect was further analyzed using a two-carrier model (equation 2) as proposed by Bansal et al.[48] and described in the *Supplementary Information*. A two-carrier conduction model is justified due to the presence of both electron and hole pockets in the Fermi surface of $IrO_2$.

$$G_{xy}(B) = eB \left( \frac{c_1\mu_2 - c_2}{(\mu_2/\mu_1 - 1)(1 + \mu_1^2 B^2)} + \frac{c_1\mu_1 - c_2}{(\mu_2/\mu_1 - 1)(1 + \mu_2^2 B^2)} \right) \qquad (2)$$

As shown in the inset of Fig. 3a, excellent fits were obtained using equation 2 with $c_1$ and $c_2$ as the fit parameters. $c_1$ and $c_2$ are a function of the carrier concentrations and mobilities of the two conduction channels. Strikingly, the two-carrier analysis revealed the presence of high mobility minority electron carriers along with low mobility majority hole carriers. As shown in Fig. 3b and 3c, the minority electron concentration varied between $10^{15}$ - $10^{17}$ cm$^{-3}$ and their mobilities exceeded 3000 cm$^2$/Vs. This extracted high mobility is quite remarkable given the traditionally low electron mobilities in most oxide materials.

Although a non-linear Hall effect has not been observed before in IrO$_2$, the presence of the different carrier types has been experimentally demonstrated[34] before using Hall effect measurements on IrO$_2$ films with different epitaxial orientations. If the Fermi surface includes both electron and hole pockets, it is plausible that Hall measurements will show both electrons and holes as majority and/or minority carriers depending on the crystal orientation. To confirm this effect for the IrO$_2$ (110) films used in this study, we also studied epitaxial IrO$_2$ (101) and IrO$_2$ (001) films on corresponding TiO$_2$ substrates. As shown in Fig. 4a, we observed that for similar film thickness, the (110) oriented films showed considerably higher resistivity as compared to the other orientations. This observation deviates from previous reports[24, 34] where the (110) oriented films were either found to have similar or lower resistivity. The origin of this discrepancy is currently not understood. Differences in the exact strain states, dislocations and point defects can be possible factors. Irrespective of these differences, the Hall effect measurements shown in Fig. 4b for the three orientations show varying dominant carrier type, in agreement with ref.[34]. We observed dominant hole conduction in the (110) oriented film and dominant electron conduction in the (101) and (001) oriented films. Remarkably, we also observed strong non-linearity in the Hall conductance for the (101) oriented sample as shown in Fig 4c. Unlike IrO$_2$ (110) film, the IrO$_2$ (101) film showed low-mobility majority electron carriers ($n_e$ = 2.23 × 10$^{22}$ cm$^{-3}$; $\mu_e$ = 8.31 cm$^2$/Vs) and high mobility minority hole carriers ($n_h$ = 3.72 × 10$^{16}$ cm$^{-3}$; $\mu_h$ = 2540 cm$^2$/Vs) as extracted from the two-carrier model. The observation of both high mobility electron and hole carriers as function of film orientation (which also governs epitaxial strain and current direction) in IrO$_2$ raises questions about their origin and necessitates further investigation. In addition, we also observed weak non-linearity starting to appear at magnetic fields higher than 6-7 T for the IrO$_2$ (001) film which needs to be further verified with Hall measurements at magnetic fields higher than 9 T.

**Conclusion:**

To summarize, we have studied the electrical transport properties in IrO$_2$ films using a combination of temperature dependent DC resistivity measurements and low temperature Hall resistance measurements. The analysis of the temperature-dependent resistivity measurements for epitaxial IrO$_2$ (110) films with different thickness showed that the interband acoustic phonon scattering and optical phonon scattering are the dominant scattering mechanisms in the normal state transport, with a weak contribution from electron-electron scattering at low temperatures. The extracted electron-phonon coupling constants agree well with previously reported values in literature and the extracted Debye temperature of 430 K agrees well with the approach to $T$-linear behavior observed around 400 K. The measured $IrO_2$ (110) samples also showed a thickness dependent in-plane resistance anisotropy which is likely a result of strain relaxation induced changes in orbital hybridization and Fermi level density of states. The effect of thickness and epitaxial strain was also observed in a previously unreported emergent non-linearity in the Hall resistance and a two-carrier model analysis of the non-linearity revealed the presence of high mobility minority carriers. The observation of multiple carrier types was also confirmed using $IrO_2$ films with different epitaxial orientations which showed varying carrier properties in terms of charge, concentration and mobilities, attesting to the complex Fermi surface and potential origins of these unusual transport features in the topological nature of the electronic band structure of $IrO_2$.

**Supporting Information:**

Experimental methods and details of scattering terms included in equation 1, extracted fit parameters for equation 1, additional analysis of resistivity and Hall conductance along with a description of the two-carrier model used to fit Hall conductance data.

**Author contributions:**

S. N. and B. J. designed the experiments. S. N. grew the $IrO_2$ films. S. N. and Z. Y. performed electrical and magneto-transport measurements and data analysis. S.N. and K.N. discussed the transport data. S. N. and B. J. wrote the manuscript. B. J. supervised all aspects of the project. All authors contributed to the discussion of the manuscript.

**Acknowledgements:**

This work was supported primarily by the Air Force Office of Scientific Research (AFOSR) through Grant Nos. FA9550-21-1-0025, and FA9550-23-1-0247. K.S. acknowledges support from the AFOSR through the Grant No. FA9550-22-1-0205. Film growth was performed using instrumentation funded by AFOSR DURIP awards FA9550-18-1-0294 and FA9550-23-1-0085.

Film growth (S.N.) was supported by the U.S. Department of Energy through award No. DE-SC0024710. Z. Y. received partial support from the UMN MRSEC program under Award No. DMR-2011401. Parts of this work were carried out at the Characterization Facility, University of Minnesota, which receives partial support from the NSF through the MRSEC program under award DMR-2011401. Device fabrication was carried out at the Minnesota Nano Center, which is supported by the NSF through the National Nano Coordinated Infrastructure under award ECCS-2025124.

**Notes:**

The authors declare no competing financial interest.

**Data availability:**

All relevant data are included in the main text and supplementary information. Any additional information can be requested from the corresponding authors.

**References:**

1. Zylbersztejn, A.; Mott, N. F., Metal-insulator transition in vanadium dioxide. *Physical Review B* **1975,** *11* (11), 4383-4395.
2. Aetukuri, N. B.; Gray, A. X.; Drouard, M.; Cossale, M.; Gao, L.; Reid, A. H.; Kukreja, R.; Ohldag, H.; Jenkins, C. A.; Arenholz, E.; Roche, K. P.; Dürr, H. A.; Samant, M. G.; Parkin, S. S. P., Control of the metal–insulator transition in vanadium dioxide by modifying orbital occupancy. *Nature Physics* **2013,** *9* (10), 661-666.
3. Taha, M.; Walia, S.; Ahmed, T.; Headland, D.; Withayachumnankul, W.; Sriram, S.; Bhaskaran, M., Insulator–metal transition in substrate-independent VO2 thin film for phase-change devices. *Scientific Reports* **2017,** *7* (1), 17899.
4. Martens, K.; Jeong, J. W.; Aetukuri, N.; Rettner, C.; Shukla, N.; Freeman, E.; Esfahani, D. N.; Peeters, F. M.; Topuria, T.; Rice, P. M.; Volodin, A.; Douhard, B.; Vandervorst, W.; Samant, M. G.; Datta, S.; Parkin, S. S. P., Field Effect and Strongly Localized Carriers in the Metal-Insulator Transition Material ${\mathrm{VO}}_{2}$. *Physical Review Letters* **2015,** *115* (19), 196401.
5. Schofield, P.; Bradicich, A.; Gurrola, R. M.; Zhang, Y.; Brown, T. D.; Pharr, M.; Shamberger, P. J.; Banerjee, S., Harnessing the Metal–Insulator Transition of VO2 in Neuromorphic Computing. *Advanced Materials* **2023,** *35* (37), 2205294.
6. Ruf, J. P.; Paik, H.; Schreiber, N. J.; Nair, H. P.; Miao, L.; Kawasaki, J. K.; Nelson, J. N.; Faeth, B. D.; Lee, Y.; Goodge, B. H.; Pamuk, B.; Fennie, C. J.; Kourkoutis, L. F.; Schlom, D. G.; Shen, K. M., Strain-stabilized superconductivity. *Nature Communications* **2021,** *12* (1), 59.
7. Jovic, V.; Koch, R. J.; Panda, S. K.; Berger, H.; Bugnon, P.; Magrez, A.; Smith, K. E.; Biermann, S.; Jozwiak, C.; Bostwick, A.; Rotenberg, E.; Moser, S., Dirac nodal lines and flat-band surface state in the functional oxide ${\mathrm{RuO}}_{2}$. *Physical Review B* **2018,** *98* (24), 241101.
8. Feng, Z.; Zhou, X.; Šmejkal, L.; Wu, L.; Zhu, Z.; Guo, H.; González-Hernández, R.; Wang, X.; Yan, H.; Qin, P.; Zhang, X.; Wu, H.; Chen, H.; Meng, Z.; Liu, L.; Xia, Z.; Sinova, J.; Jungwirth, T.; Liu, Z., An anomalous Hall effect in altermagnetic ruthenium dioxide. *Nature Electronics* **2022,** *5* (11), 735-743.
9. Šmejkal, L.; Sinova, J.; Jungwirth, T., Emerging Research Landscape of Altermagnetism. *Physical Review X* **2022,** *12* (4), 040501.
10. Berlijn, T.; Snijders, P. C.; Delaire, O.; Zhou, H. D.; Maier, T. A.; Cao, H. B.; Chi, S. X.; Matsuda, M.; Wang, Y.; Koehler, M. R.; Kent, P. R. C.; Weitering, H. H., Itinerant Antiferromagnetism in ${\mathrm{RuO}}_{2}$. *Physical Review Letters* **2017,** *118* (7), 077201.
11. Jeong, S. G.; Choi, I. H.; Nair, S.; Buiarelli, L.; Pourbahari, B.; Oh, J. Y.; Bassim, N.; Seo, A.; Choi, W. S.; Fernandes, R. M.; Birol, T.; Zhao, L.; Lee, J. S.; Jalan, B., Altermagnetic polar metallic phase in ultra-thin epitaxially-strained RuO$_2$ films. *<http://arxiv.org/abs/2405.05838>* **2024**.
12. Bose, A.; Schreiber, N. J.; Jain, R.; Shao, D.-F.; Nair, H. P.; Sun, J.; Zhang, X. S.; Muller, D. A.; Tsymbal, E. Y.; Schlom, D. G.; Ralph, D. C., Author Correction: Tilted spin current generated by the collinear antiferromagnet ruthenium dioxide. *Nature Electronics* **2022,** *5* (10), 706-706.
13. Jiang, Y.-Y.; Wang, Z.-A.; Samanta, K.; Zhang, S.-H.; Xiao, R.-C.; Lu, W. J.; Sun, Y. P.; Tsymbal, E. Y.; Shao, D.-F., Prediction of giant tunneling magnetoresistance in $\mathrm{Ru}{\mathrm{O}}_{2}/\mathrm{Ti}{\mathrm{O}}_{2}/\mathrm{Ru}{\mathrm{O}}_{2}$ (110) antiferromagnetic tunnel junctions. *Physical Review B* **2023,** *108* (17), 174439.
14. Shao, D.-F.; Zhang, S.-H.; Li, M.; Eom, C.-B.; Tsymbal, E. Y., Spin-neutral currents for spintronics. *Nature Communications* **2021,** *12* (1), 7061.
15. Zhang, T. X.; Coughlin, A. L.; Lu, C.-K.; Heremans, J. J.; Zhang, S. X., Recent progress on topological semimetal IrO2: electronic structures, synthesis, and transport properties. *Journal of Physics: Condensed Matter* **2024,** *36* (27), 273001.
16. Pesin, D.; Balents, L., Mott physics and band topology in materials with strong spin–orbit interaction. *Nature Physics* **2010,** *6* (5), 376-381.
17. Laurell, P.; Fiete, G. A., Topological Magnon Bands and Unconventional Superconductivity in Pyrochlore Iridate Thin Films. *Physical Review Letters* **2017,** *118* (17), 177201.
18. Savary, L.; Moon, E.-G.; Balents, L., New Type of Quantum Criticality in the Pyrochlore Iridates. *Physical Review X* **2014,** *4* (4), 041027.
19. Wang, F.; Senthil, T., Twisted Hubbard Model for ${\mathrm{Sr}}_{2}{\mathrm{IrO}}_{4}$: Magnetism and Possible High Temperature Superconductivity. *Physical Review Letters* **2011,** *106* (13), 136402.
20. Chen, Y.; Lu, Y.-M.; Kee, H.-Y., Topological crystalline metal in orthorhombic perovskite iridates. *Nature Communications* **2015,** *6* (1), 6593.
21. Liu, Z. T.; Li, M. Y.; Li, Q. F.; Liu, J. S.; Li, W.; Yang, H. F.; Yao, Q.; Fan, C. C.; Wan, X. G.; Wang, Z.; Shen, D. W., Direct observation of the Dirac nodes lifting in semimetallic perovskite SrIrO3 thin films. *Scientific Reports* **2016,** *6* (1), 30309.
22. Nan, T.; Anderson, T. J.; Gibbons, J.; Hwang, K.; Campbell, N.; Zhou, H.; Dong, Y. Q.; Kim, G. Y.; Shao, D. F.; Paudel, T. R.; Reynolds, N.; Wang, X. J.; Sun, N. X.; Tsymbal, E. Y.; Choi, S. Y.; Rzchowski, M. S.; Kim, Y. B.; Ralph, D. C.; Eom, C. B., Anisotropic spin-orbit torque generation in epitaxial SrIrO3 by symmetry design. *Proceedings of the National Academy of Sciences* **2019,** *116* (33), 16186-16191.
23. Sun, Y.; Zhang, Y.; Liu, C.-X.; Felser, C.; Yan, B., Dirac nodal lines and induced spin Hall effect in metallic rutile oxides. *Physical Review B* **2017,** *95* (23), 235104.
24. Nelson, J. N.; Ruf, J. P.; Lee, Y.; Zeledon, C.; Kawasaki, J. K.; Moser, S.; Jozwiak, C.; Rotenberg, E.; Bostwick, A.; Schlom, D. G.; Shen, K. M.; Moreschini, L., Dirac nodal lines protected against spin-orbit interaction in ${\mathrm{IrO}}_{2}$. *Physical Review Materials* **2019,** *3* (6), 064205.
25. Xu, X.; Jiang, J.; Shi, W. J.; Süß, V.; Shekhar, C.; Sun, S. C.; Chen, Y. J.; Mo, S. K.; Felser, C.; Yan, B. H.; Yang, H. F.; Liu, Z. K.; Sun, Y.; Yang, L. X.; Chen, Y. L., Strong spin-orbit coupling and Dirac nodal lines in the three-dimensional electronic structure of metallic rutile ${\mathrm{IrO}}_{2}$. *Physical Review B* **2019,** *99* (19), 195106.
26. Das, P. K.; Sławińska, J.; Vobornik, I.; Fujii, J.; Regoutz, A.; Kahk, J. M.; Scanlon, D. O.; Morgan, B. J.; McGuinness, C.; Plekhanov, E.; Di Sante, D.; Huang, Y.-S.; Chen, R.-S.; Rossi, G.; Picozzi, S.; Branford, W. R.; Panaccione, G.; Payne, D. J., Role of spin-orbit coupling in the electronic structure of $\mathrm{Ir}{\mathrm{O}}_{2}$. *Physical Review Materials* **2018,** *2* (6), 065001.
27. Bose, A.; Nelson, J. N.; Zhang, X. S.; Jadaun, P.; Jain, R.; Schlom, D. G.; Ralph, D. C.; Muller, D. A.; Shen, K. M.; Buhrman, R. A., Effects of Anisotropic Strain on Spin–Orbit Torque Produced by the Dirac Nodal Line Semimetal IrO2. *ACS Applied Materials & Interfaces* **2020,** *12* (49), 55411-55416.

28. Patton, M.; Gurung, G.; Shao, D.-F.; Noh, G.; Mittelstaedt, J. A.; Mazur, M.; Kim, J.-W.; Ryan, P. J.; Tsymbal, E. Y.; Choi, S.-Y.; Ralph, D. C.; Rzchowski, M. S.; Nan, T.; Eom, C.-B., Symmetry Control of Unconventional Spin–Orbit Torques in IrO2. *Advanced Materials* **2023,** *35* (39), 2301608.
29. Kahk, J. M.; Poll, C. G.; Oropeza, F. E.; Ablett, J. M.; Céolin, D.; Rueff, J. P.; Agrestini, S.; Utsumi, Y.; Tsuei, K. D.; Liao, Y. F.; Borgatti, F.; Panaccione, G.; Regoutz, A.; Egdell, R. G.; Morgan, B. J.; Scanlon, D. O.; Payne, D. J., Understanding the Electronic Structure of ${\mathrm{IrO}}_{2}$ Using Hard-X-ray Photoelectron Spectroscopy and Density-Functional Theory. *Physical Review Letters* **2014,** *112* (11), 117601.
30. Kim, W. J.; Kim, S. Y.; Kim, C. H.; Sohn, C. H.; Korneta, O. B.; Chae, S. C.; Noh, T. W., Spin-orbit coupling induced band structure change and orbital character of epitaxial $\mathrm{Ir}{\mathrm{O}}_{2}$ films. *Physical Review B* **2016,** *93* (4), 045104.
31. Panda, S. K.; Bhowal, S.; Delin, A.; Eriksson, O.; Dasgupta, I., Effect of spin orbit coupling and Hubbard $U$ on the electronic structure of IrO${}_{2}$. *Physical Review B* **2014,** *89* (15), 155102.
32. Arias-Egido, E.; Laguna-Marco, M. A.; Piquer, C.; Jiménez-Cavero, P.; Lucas, I.; Morellón, L.; Gallego, F.; Rivera-Calzada, A.; Cabero-Piris, M.; Santamaria, J.; Fabbris, G.; Haskel, D.; Boada, R.; Díaz-Moreno, S., Dimensionality-driven metal–insulator transition in spin–orbit-coupled IrO2. *Nanoscale* **2021,** *13* (40), 17125-17135.
33. Kawasaki, J. K.; Kim, C. H.; Nelson, J. N.; Crisp, S.; Zollner, C. J.; Biegenwald, E.; Heron, J. T.; Fennie, C. J.; Schlom, D. G.; Shen, K. M., Engineering Carrier Effective Masses in Ultrathin Quantum Wells of ${\mathrm{IrO}}_{2}$. *Physical Review Letters* **2018,** *121* (17), 176802.
34. Uchida, M.; Sano, W.; Takahashi, K. S.; Koretsune, T.; Kozuka, Y.; Arita, R.; Tokura, Y.; Kawasaki, M., Field-direction control of the type of charge carriers in nonsymmorphic ${\mathrm{IrO}}_{2}$. *Physical Review B* **2015,** *91* (24), 241119.
35. Goodenough, J. B., Direct Cation- -Cation Interactions in Several Oxides. *Physical Review* **1960,** *117* (6), 1442-1451.
36. Goodenough, J. B., Metallic oxides. *Progress in Solid State Chemistry* **1971,** *5*, 145-399.
37. Occhialini, C. A.; Bisogni, V.; You, H.; Barbour, A.; Jarrige, I.; Mitchell, J. F.; Comin, R.; Pelliciari, J., Local electronic structure of rutile ${\mathrm{RuO}}_{2}$. *Physical Review Research* **2021,** *3* (3), 033214.
38. Ryden, W. D.; Lawson, A. W.; Sartain, C. C., Temperature dependence of the resistivity of RuO2 and IrO2. *Physics Letters A* **1968,** *26* (5), 209-210.
39. Nunn, W.; Manjeshwar, A. K.; Yue, J.; Rajapitamahuni, A.; Truttmann, T. K.; Jalan, B., Novel synthesis approach for "stubborn" metals and metal oxides. *Proceedings of the National Academy of Sciences* **2021,** *118* (32), e2105713118.
40. Nair, S.; Yang, Z.; Lee, D.; Guo, S.; Sadowski, J. T.; Johnson, S.; Saboor, A.; Li, Y.; Zhou, H.; Comes, R. B.; Jin, W.; Mkhoyan, K. A.; Janotti, A.; Jalan, B., Engineering metal oxidation using epitaxial strain. *Nature Nanotechnology* **2023,** *18* (9), 1005-1011.
41. Liao, Z.; Li, F.; Gao, P.; Li, L.; Guo, J.; Pan, X.; Jin, R.; Plummer, E. W.; Zhang, J., Origin of the metal-insulator transition in ultrathin films of $\mathrm{L}{\mathrm{a}}_{2/3}\mathrm{S}{\mathrm{r}}_{1/3}\mathrm{Mn}{\mathrm{O}}_{3}$. *Physical Review B* **2015,** *92* (12), 125123.

42. Wang, G.; Wang, Z.; Meng, M.; Saghayezhian, M.; Chen, L.; Chen, C.; Guo, H.; Zhu, Y.; Plummer, E. W.; Zhang, J., Role of disorder and correlations in the metal-insulator transition in ultrathin ${\mathrm{SrVO}}_{3}$ films. *Physical Review B* **2019,** *100* (15), 155114.
43. Rajapitamahuni, A. K.; Nair, S.; Yang, Z.; Kamath Manjeshwar, A.; Gyo Jeong, S.; Nunn, W.; Jalan, B., Thickness-dependent insulator-to-metal transition in epitaxial RuO2 films. *arXiv e-prints* **2023**, arXiv:2312.07869.
44. Lin, J. J.; Huang, S. M.; Lin, Y. H.; Lee, T. C.; Liu, H.; Zhang, X. X.; Chen, R. S.; Huang, Y. S., Low temperature electrical transport properties of RuO2 and IrO2 single crystals. *Journal of Physics: Condensed Matter* **2004,** *16* (45), 8035.
45. Lin, Y. H.; Sun, Y. C.; Jian, W. B.; Chang, H. M.; Huang, Y. S.; Lin, J. J., Electrical transport studies of individual IrO2 nanorods and their nanorod contacts. *Nanotechnology* **2008,** *19* (4), 045711.
46. Glassford, K. M.; Chelikowsky, J. R., Electronic and structural properties of ${\mathrm{RuO}}_{2}$. *Physical Review B* **1993,** *47* (4), 1732-1741.
47. Price, W. L. V., Extension of van der Pauw's theorem for measuring specific resistivity in discs of arbitrary shape to anisotropic media. *Journal of Physics D: Applied Physics* **1972,** *5* (6), 1127.
48. Bansal, N.; Kim, Y. S.; Brahlek, M.; Edrey, E.; Oh, S., Thickness-Independent Transport Channels in Topological Insulator ${\mathrm{Bi}}_{2}{\mathrm{Se}}_{3}$ Thin Films. *Physical Review Letters* **2012,** *109* (11), 116804.
49. Bongers, P. F., Anisotropy of the electrical conductivity of VO2 single crystals. *Solid State Communications* **1965,** *3* (9), 275-277.
50. Lewis, S. P.; Allen, P. B.; Sasaki, T., Band structure and transport properties of ${\mathrm{CrO}}_{2}$. *Physical Review B* **1997,** *55* (16), 10253-10260.

**Figure 1.**

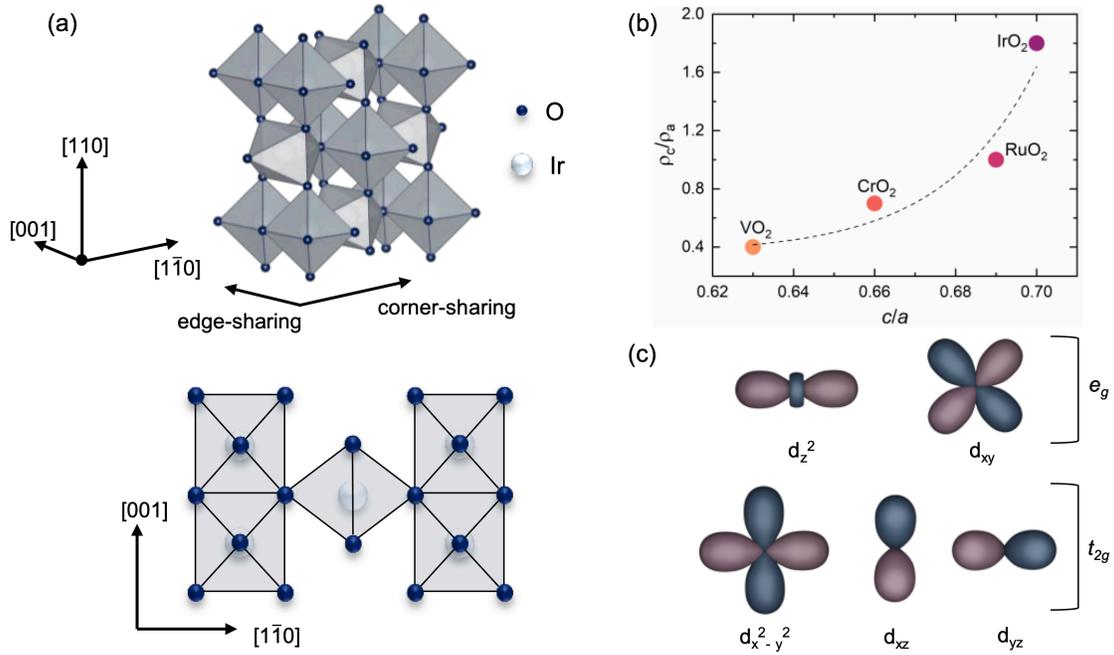

**Figure 1:** (a) (Top) Rutile IrO$_2$ crystal structure with modified unit cell for thin film grown along the [110] out of plane direction. (Bottom) 2D projection of the crystal along [110] for better visualization of the edge shared and corner shared octahedral network along the [001] and [1$\bar{1}$0] crystal directions. (b) Reported electrical resistivity anisotropy between the rutile $c$ and $a$ axis for bulk single crystals of metallic rutile VO$_2$[49], CrO$_2$[50], RuO$_2$ and IrO$_2$[38]. Dashed line is a trend line. (c) Schematic of the *d*-orbital shapes and rutile crystal field splitting into $t_{2g}$ and $e_g$ manifolds with the x and y axes defined along the [001] and [1$\bar{1}$0] directions respectively

**Figure 2.**

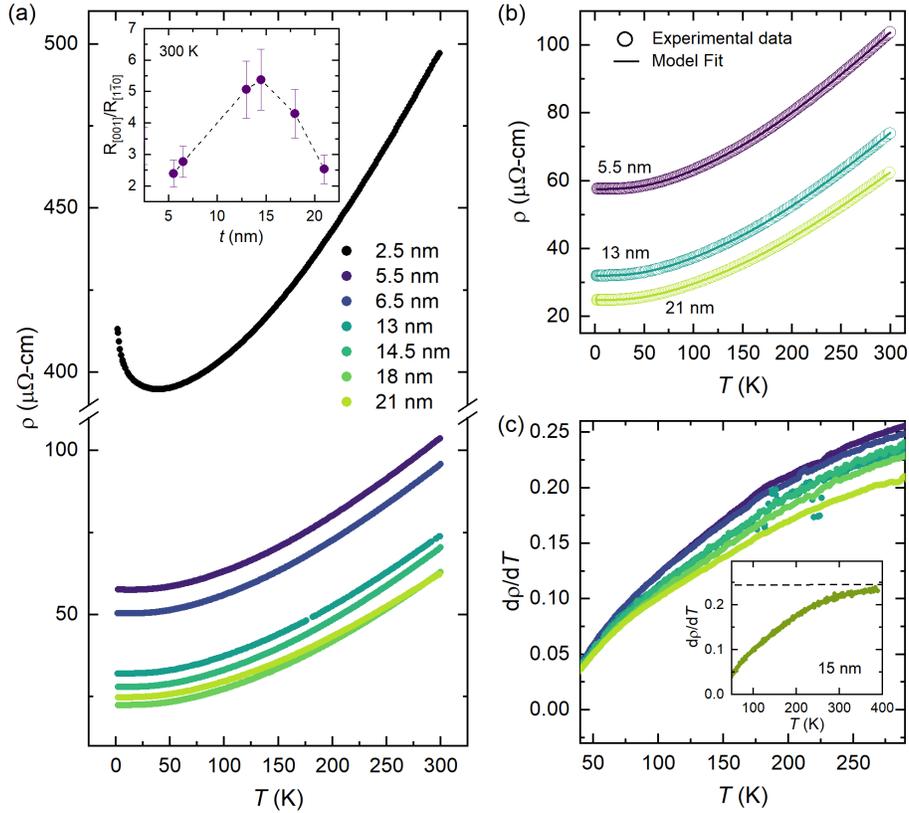

**Figure 2:** (a) Resistivity as a function of temperature for different thicknesses of $IrO_2$ (110) films grown on $TiO_2$ (110) substrate. (inset) Anisotropy in the measured resistance along the in-plane [001] and [1$\bar{1}$0] crystal directions plotted as a function of film thickness. (Error bars are calculated using an arbitrary estimate of 12.5% error in the distance between the wire-bonded contact points). (b) Measured resistivity (open circles) and model fit to equation 1 (solid lines) for 5.5 nm, 13 nm and 21 nm $IrO_2$ (110) films shown in (a). (c) First order temperature derivative of resistivity plotted as a function of temperature for the $IrO_2$ (110) films shown in (a). (inset) First order temperature derivative of resistivity plotted as a function of temperature for a representative sample measured upto 390 K showing an approach to $T$-linear behavior indicated by the black dashed line.

**Figure 3.**

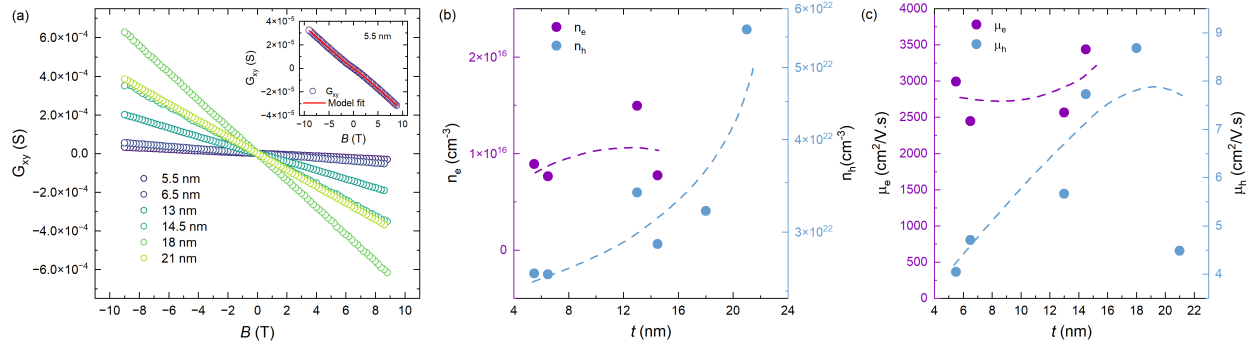

**Figure 3:** (a) Hall conductance as a function of magnetic field at 1.8K for different thicknesses of $IrO_2$ (110) films grown on $TiO_2$ (110) substrates. (inset) Hall conductance as a function of magnetic field for a 5.5 nm $IrO_2$ (110) film for better visualization of non-linear Hall effect along with two-carrier model fit. Extracted (b) carrier concentrations and (c) mobilities for the electron and hole channels from two-carrier model fits plotted as a function of film thickness. Dashed lines are guidelines. Straight line fits were used for 18 nm and 21 nm films due to vanishing non-linearity and hence, solely hole conduction.

**Figure 4.**

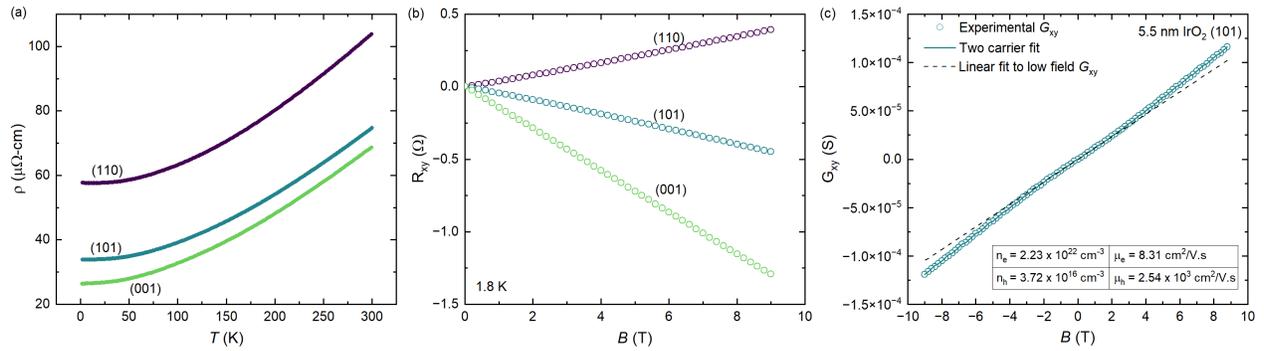

**Figure 4:** (a) Resistivity as a function of temperature for $IrO_2$ films of nominally same thickness (~ 5.5-5.8 nm) grown along different out of plane orientations on corresponding $TiO_2$ substrates. (b) Hall resistance as a function of magnetic field at 1.8 K for $IrO_2$ films shown in (a). (c) Hall conductance as a function of magnetic field for $IrO_2$ (101) film along with two-carrier model fit used to extract the carrier concentrations and mobilities of the two channels shown in the inset table. A straight line fit to the low magnetic field (-0.4 T to 0.4 T) Hall conductance is also plotted for better visualization of the non-linear Hall effect.